\documentclass[aps,prb,twocolumn,10pt, showkeys,showpacs,amsmath,amsfonts,floatfix,superscriptaddress,nofootinbib,longbibliography]{revtex4-2}

\usepackage{silence}
 \ErrorsOff*
\usepackage{graphicx} 
\graphicspath{{figures/}}
\usepackage{dcolumn} 
\usepackage{color}
\usepackage{amsmath} 
\usepackage{blindtext, lipsum}
\usepackage{braket}
\usepackage[breaklinks, colorlinks, bookmarks=false, citecolor=magenta,
            linkcolor=cyan, urlcolor=magenta]{hyperref}
\usepackage{cleveref} 
\usepackage{changes}

\newcommand{\sgn}[0]{\mathrm{sgn}}

\newcommand\norm[1]{\left\lVert#1\right\rVert}

\begin{document}

\title{Krylov space dynamics of ergodic and dynamically frozen Floquet systems}

\author{Luke Staszewski}
\email{lstaszewski@pks.mpg.de} 
\affiliation{Max Planck Institute for the Physics of Complex Systems, N\"othnitzer Strasse 38, Dresden 01187, Germany}

\author{Asmi Haldar}
\affiliation{Laboratoire de Physique Théorique - IRSAMC Paul Sabatier University, 118, Route de Narbonne building 3R1B4 31400 Toulouse, France}

\author{Pieter W. Claeys}
\affiliation{Max Planck Institute for the Physics of Complex Systems, N\"othnitzer Strasse 38, Dresden 01187, Germany}

\author{Alexander Wietek}
\email{awietek@pks.mpg.de}
\affiliation{Max Planck Institute for the Physics of Complex Systems, N\"othnitzer Strasse 38, Dresden 01187, Germany}

\begin{abstract}
In isolated quantum many-body systems periodically driven in time, the
asymptotic dynamics at late times can exhibit distinct behavior such as
thermalization or dynamical freezing. Understanding the properties of and the
convergence towards infinite-time (nonequilibrium) steady states however remains
a challenging endeavor. We propose a physically motivated Krylov space
perspective on Floquet thermalization which offers a natural framework to study
rates of convergence towards steady states and, simultaneously, an efficient
numerical algorithm to evaluate infinite-time averages of observables within the
diagonal ensemble. The effectiveness of our algorithm is demonstrated by
applying it to the periodically driven mixed-field Ising model, reaching system
sizes of up to 30 spins. Our method successfully resolves the transition between
the ergodic and dynamically frozen phases and provides insight into the nature
of the Floquet eigenstates across the phase diagram. Furthermore, we show that
the long-time behavior is encoded within the localization properties of the Ritz
vectors under the Floquet evolution, providing an accurate diagnostic of
ergodicity.
\end{abstract}

\date{\today}

\maketitle

\section{Introduction}
\label{sec:intro}

Understanding the late-time dynamics of isolated quantum systems remains a
central problem at the forefront of modern physics. In periodically driven
systems, such dynamics are generally expected to follow the prescriptions of the
Floquet eigenstate thermalization hypothesis (Floquet
ETH)~\cite{Srednicki_DE,dalessio_quantum_2016,Lazarides_Das_Moessner,Rigol_Infinite_T},
describing the drive-induced heating towards an infinite-temperature
state~\cite{bukov_heating_2016,weidinger_floquet_2017,machado_exponentially_2019,rubio-abadal_floquet_2020,ikeda_fermis_2021}.
Identifying regimes where such ergodicity is avoided opens access to rich and
unconventional
physics~\cite{bukov_universal_2015,moessner_equilibration_2017,oka_floquet_2019,rudner_band_2020,khemani_brief_2019,schindler_geometric_2025,Moudgalya_HSF}.
However, numerically probing and characterizing these regimes remains a
formidable challenge. Distinguishing true ergodicity breaking from
prethermalization, where ergodicity may only set in at large but finite time
scales that mimic genuinely diverging ones, requires both large system sizes and
late times (as exemplified in strongly disordered
systems~\cite{Abanin_challenges_2021}).

\begin{figure}[!t]
    \centering
    \includegraphics[width=\linewidth]{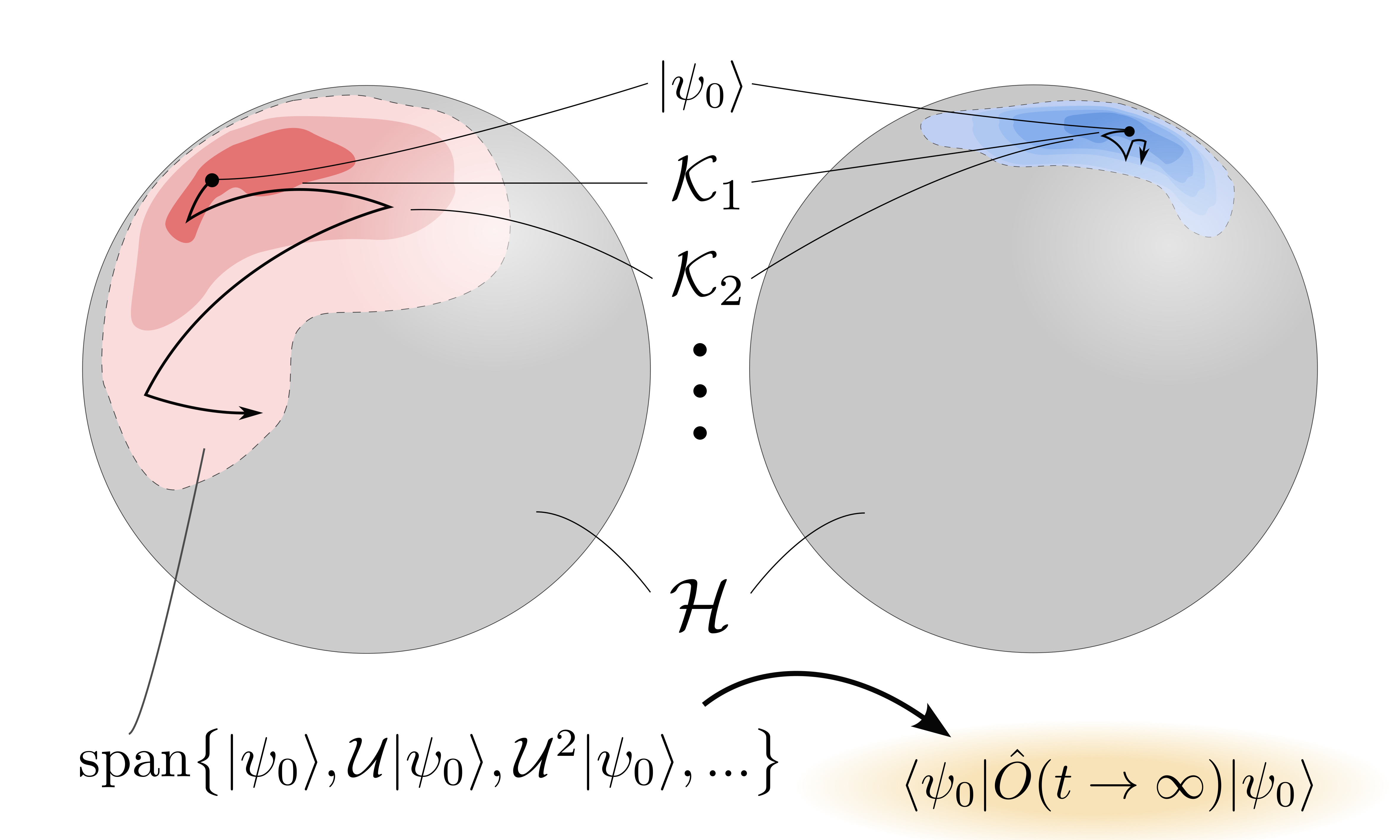}
    \caption{ We introduce a Krylov space approximation for computing the
    infinite-time average (“DEA”) of observables in periodically driven systems.
    The figure illustrates two contrasting scenarios: (left) an ergodic regime,
    where observables rapidly relax to equilibrium, and (right) a dynamically
    frozen regime, where emergent conservation laws keep observables locked near
    their initial values for long times. Our Krylov subspace approach yields
    accurate infinite-time averages in both cases. }
    \label{fig:schematic}
\end{figure}

In isolated, periodically driven systems, the time evolution over a single
period is described by the unitary Floquet operator $\mathcal{U}$. Its
eigendecomposition,
\begin{equation}
    \label{eq:floquetoperator}
    \mathcal{U}\ket{\mu_i} = \xi_i \ket{\mu_i},
\end{equation}
defines the Floquet eigenstates $\ket{\mu_i}$ and eigenvalues $\xi_i \in S^1$.
These eigenstates play a central role in diagnosing ergodicity breaking in
larger systems within feasible computational effort. Ergodicity breaking
manifests as freezing or localization around low-entanglement initial states,
such as quantum scars~\cite{Scar_Lukin_Nature, Scar_Abanin_NatPhys,
Scar_Shiraishi-Mori, Scar_Abanin_PRB, Scar_Vedika_PRB, Scar_Lukin_PRL,
Scar_Serbyn_PRL, Scar_Bernevig_1, Scar_Vedika_Rahul, Scar_Bernevig_2},
Hilbert-space
fragmentation~\cite{Sala_HSF_2019,Moudgalya_HSF,Wang2025_HSF_Superconducting},
many-body localization~\cite{Altshuler_MBL,ADP_MBL}, or dynamically frozen
({DF})~\cite{AD-DMF, AD-SDG, Mahesh_Freezing, Russomanno_Dynamical_Freezing,
Kris-Periodic, Analabha_Dynamical_Freezing, Naveen_Dynamical_Freezing,Onset,
Asmi_DF_PRX_2021,Bhaskar_DF,Diptiman_DF, Analabha_Mori_Rehman_DF,
Debanjan_DF_QDot, Debanjan_DF_Transmon,Krishanu_DF,Asmi_Flq_Rev,
Tista_KS_Rev,AH_DF_TDL} systems stabilized by emergent conserved quantities.

Given an initial state $\ket{\psi_0}$, the late-time non-oscillatory dynamics of
an observable $\mathcal{O}$ with respect to the Floquet time evolution
$\mathcal{U}^m\ket{\psi_0}$ for $m\!\to\!\infty$ is captured by the diagonal
ensemble average
(DEA)~\cite{Marcos_Nature_2008,Marcos_PRL_2013,Rigol_Infinite_T},
\begin{equation}
\label{eq:dea}
\braket{ \mathcal{O} }_{\mathrm{DEA}} = \sum_{i=1}^{D}
{|\langle \mu_i | \psi_0 \rangle |}^{2}
\braket{\mu_{i} | \mathcal{O} | \mu_{i}} ,
\end{equation}
where $D$ denotes the Hilbert space dimension. Deviations of this DEA from the
ergodic value, $\mathrm{Tr}[\mathcal{O}]/D$, directly indicate the breaking of
ergodicity. For local operators $\mathcal{O}$, the DEA reveals the system's fate
in the long-time limit, enabling conclusive statements about its ergodicity,
thermalization, and final ensemble description, bridging quantum dynamics with
statistical
mechanics~\cite{Rigol_Infinite_T,Rigol_GGE_1,Reimann,dalessio_quantum_2016,Ponte_2015}.
A direct computation of the DEA however requires all eigenstates of the relevant
Hamiltonian, making it exponentially costly and limiting analyses to small
systems. This bottleneck similarly appears in full time-evolution simulations,
where capturing late-time quantities or linear entanglement growth rapidly
becomes intractable.

Yet, the dynamics of local observables are often far simpler than this
exponential complexity suggests. In ergodic systems, $\braket{\mathcal{O}(t)}$
typically relaxes to its thermal value after only modest evolution times. In
nonergodic systems, emergent conservation laws can cause observables to remain
effectively
frozen~\cite{Asmi_DF_PRX_2021,Bhaskar_DF,Diptiman_DF,Tista_KS_Rev,Debanjan_DF_QDot,Debanjan_DF_Transmon,lu_NV_exp}.
In both limits, full diagonalization appears excessive. Motivated by these
observations, we propose a framework for tackling ergodicity breaking and the
approach to the DEA in terms of a Krylov space approximation. Using this
approach we introduce an efficient Arnoldi-based algorithm for evaluating the
DEA. We demonstrate its effectiveness using a driven transverse field Ising
model, exhibiting a transition from an ergodic to a dynamically frozen regime,
depicted in \cref{fig:schematic}. Finally, we demonstrate how Ritz vector
localization can serve as a reliable method for studying ergodicity.

\section{Krylov approximation of the DEA}

Within the first $m-1$ cycles (discrete time steps) of the Floquet evolution,
the system explores the space spanned by the states $\mathcal{U}^n\ket{\psi_0},
n=0,\ldots,m-1$. These states define the $m$-th Krylov space,
\begin{equation}
    \mathcal{K}_m (\mathcal{U}, |\psi_0\rangle ) = 
    \text{span}\left\{ \ket{\psi_0}, \mathcal{U}\ket{\psi_0}, \ldots, \mathcal{U}^{(m-1)}\ket{\psi_0} \right\}.
\label{eq:krylovspace}
\end{equation}
In the following, we abbreviate $\mathcal{K}_m \equiv \mathcal{K}_m(\mathcal{U},
|\psi_0\rangle)$. We denote by $\mathcal{P}_m = V_m V_m^\dagger$ the projector
onto $\mathcal{K}_m$, where $V_m$ denotes a matrix whose columns form an
orthonormal basis of $\mathcal{K}_m$. Moreover, we define the projected Floquet
operator as, 
\begin{equation}
\label{eq:projectedfloquet}
\tilde{\mathcal{U}}_m = \mathcal{P}_m^\dagger \mathcal{U}\mathcal{P}_m.
\end{equation}
We now introduce the Krylov-space projected DEA as 
\begin{equation}
\label{eq:dea_krylov}
\braket{ \mathcal{O} }_{\mathrm{DEA},m} = \sum_{i=1}^{m}
{|\langle \tilde{\mu}_i | \psi_0 \rangle|}^{2}
\braket{\tilde{\mu}_{i} | \mathcal{O} | \tilde{\mu}_{i}}/\mathcal{N}.
\end{equation}
Here the $\ket{\tilde{\mu}_i}$ denote the eigenvectors of the projected Floquet
operator (also referred to as Ritz vectors),
\begin{equation}
    \label{eq:floquetoperatorprojectedeigenvalues}
    \tilde{\mathcal{U}}_m\ket{\tilde{\mu}_i} = \tilde{\xi}_i \ket{\tilde{\mu}_i},
\end{equation}
for which $\mathcal{P}_m\ket{\tilde{\mu}_i} \neq 0$. While the diagonalization
of a unitary matrix gives rise to an orthonormal basis, the projected Floquet
operator is not necessarily unitary, which can lead to a loss of orthogonality
of the generated Ritz vectors. We include a normalization factor given by
\begin{equation}
\label{eq:dea_krylov_norm}
\mathcal{N} = \sum_{i=1}^{m}
{|\langle \tilde{\mu}_i | \psi_0 \rangle|}^{2}
\end{equation}
in \cref{eq:dea_krylov} to remedy this loss of orthogonality. We discuss other
options such as the use of left and right eigenvectors or the isometric Arnoldi
routine~\cite{arnoldi_isometric} in \cref{appendix:algorithm_choices}. The
projected DEA $\braket{ \mathcal{O} }_{\mathrm{DEA},m} $ in \cref{eq:dea_krylov}
describes the long-time limit of the Floquet dynamics restricted to the $m$-th
Krylov space $\mathcal{K}_m$. Thus, $\braket{ \mathcal{O} }_{\mathrm{DEA},m}$ is
independent of the choice of basis of $\mathcal{K}_m$.  The deviation from the
exact result, 
\begin{equation}
    \label{eq:deviation}
    \epsilon_m = |\braket{ \mathcal{O} }_{\mathrm{DEA}} - \braket{ \mathcal{O} }_{\mathrm{DEA, m}}|,
\end{equation}
is a natural measure of the proximity of the system to the infinite-time
(nonequilibrium) steady state after $m-1$ Floquet cycles. In other words,
$\epsilon_m$ determines the accuracy to which the Krylov space $\mathcal{K}_m$
captures the infinite-time dynamics of the operator $\mathcal{O}$ as illustrated
in \cref{fig:schematic}, where typically $m \ll D$. We propose an iterative
Arnoldi-based algorithm for evaluating $\braket{ \mathcal{O} }_{\mathrm{DEA},m}$
in \cref{appendix:algorithm}. Throughout the text, we refer to the first $m$
iterations of the algorithm as corresponding to the Krylov subspace of dimension
$m$.

\section{Numerical Results}
\label{sec:results}

We demonstrate the effectiveness of the algorithm in the context of dynamical
freezing (DF)~\cite{AD-DMF, Onset}. Dynamical freezing is a phenomenon where a
periodically driven quantum many-body system evades ergodicity due to the
emergence of conserved local operators (ECO) under strong driving. These
conservation laws are approximate, yet perpetual. DF has been studied in systems
mappable to free fermions and hard-core bosons~\cite{AD-DMF, AD-SDG,
Mahesh_Freezing, Russomanno_Dynamical_Freezing, Kris-Periodic,
Analabha_Dynamical_Freezing, Naveen_Dynamical_Freezing}, as well as in
interacting systems~\cite{Onset, Asmi_DF_PRX_2021,Bhaskar_DF,Diptiman_DF,
Analabha_Mori_Rehman_DF, Debanjan_DF_QDot,
Debanjan_DF_Transmon,Krishanu_DF,Asmi_Flq_Rev, Tista_KS_Rev}. However, studies
of DEAs for the latter have remained restricted to small system sizes.  


We consider an interacting Floquet system with the following time-dependent
Hamiltonian, describing a uniform chain of interacting Ising spins where the
longitudinal field is periodically switched:
%

\begin{figure}[!t]
    \centering
    \includegraphics[width=0.98\linewidth]{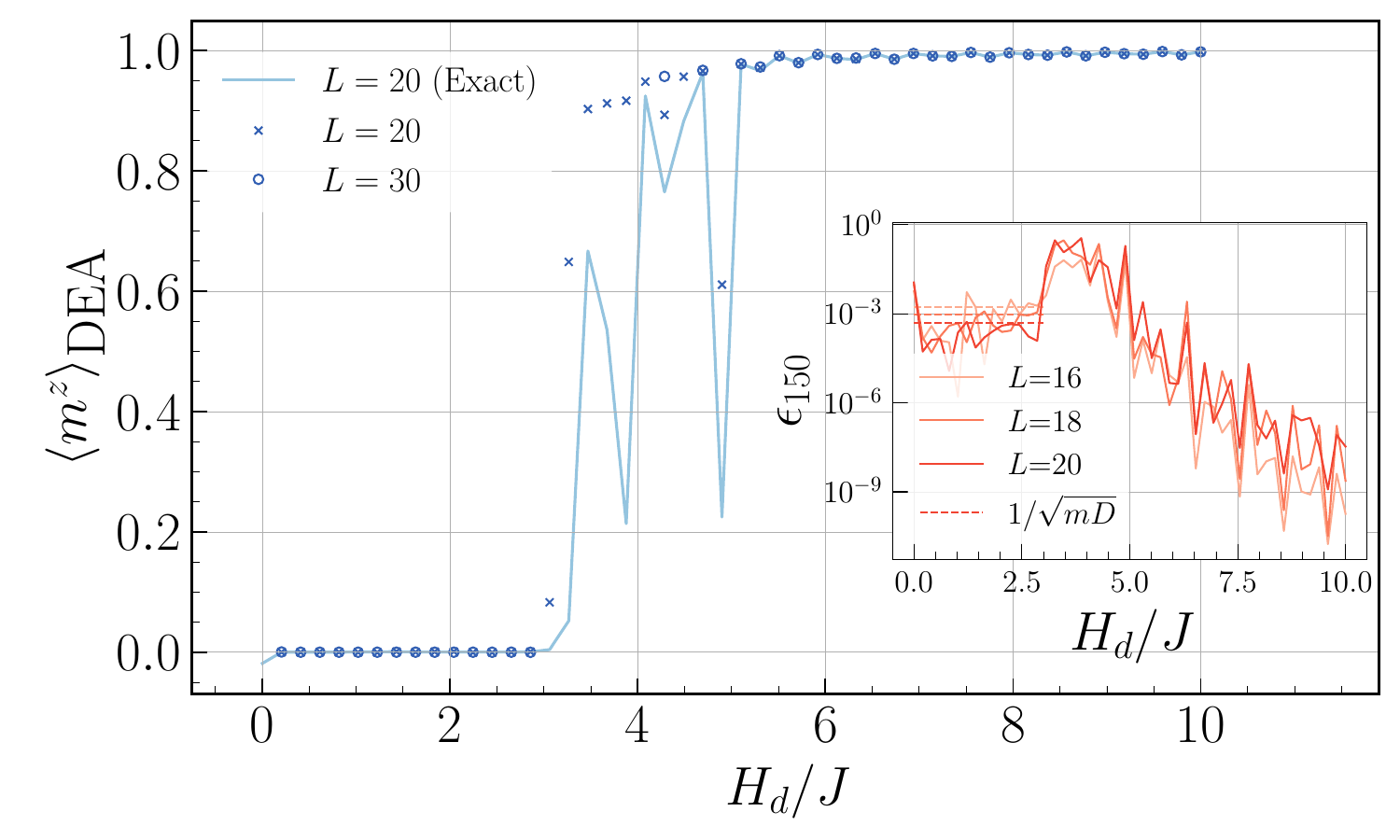}
    \caption{ Infinite-time average of the total magnetization density, $m_z$,
        showing the transition to a dynamically frozen regime at large $H_d/J$.
        The solid line shows the exact long-time average for a 20-site chain.
        Crosses denote the same quantity computed with our algorithm after 150
        iterations also for the same system size and circles mark converged
        results for a 30-site chain. Inset shows the error $\epsilon_{150}$
        after 150 iterations of our Krylov algorithm with lighter to darker
        showing increasing system sizes. Dashed lines show a suggestive scaling
        of the error in the ergodic regime. }

    \label{fig:dea}
    
\end{figure}

\begin{align}
\label{eq:tfi-hamil}
    H(t) =& -~\sum_{i=1}^L   \bigg( 
    J \sigma_i^z \sigma_{i+1}^z +  \kappa 
    \sigma_i^z \sigma_{i+2}^z + h^z \sigma_i^z  
    + h^x \sigma^x_i \Bigg) \nonumber\\ 
   & \qquad + H_d \; \sgn \left(\sin   \omega t  \right) \Bigg(\sum_{i=1}^L \sigma_i^z\Bigg),
\end{align}
where $\sigma_i^{x,y,z}$ are Pauli matrices. Following Ref.~\cite{AH_DF_TDL}, we
set $\kappa=-0.25J$, $h^z=-0.075J$, $h^x=0.577J$, use periodic boundaries and
fix the drive period $T=4\pi/J$. Varying the drive amplitude $H_d/J$ tunes the
system between ergodic and dynamically frozen regimes. We take a fully polarized
initial state $\ket{\psi_0}=\ket{\uparrow\cdots\uparrow}_z$ and compute the
long-time average of the total magnetization density, $m^z
=\frac{1}{L}\sum_{i}\sigma^z_i $. A vanishing DEA, $\braket{m^z}_{\rm DEA}
\approx 0$, indicates ergodicity, whereas a nonzero DEA close to the initial
(maximal) expectation value, $\braket{m^z}_{\rm DEA} \approx 1$, is a signature
of dynamical freezing.

\Cref{fig:dea} shows the results of our Krylov algorithm for system sizes up to
$L=30$, exceeding the reach of exact diagonalization, along with comparisons to
exact results for $L=20$. The latter were obtained after 150 iterations of the
algorithm, with the corresponding error $\epsilon_{150}$ shown in the inset for
various system sizes. For $L=30$, we only show the results where the absolute
difference of the last two iterations has converged to less than $10^{-4}$. The
algorithm efficiently reproduces both ergodic relaxation and dynamical freezing.

\begin{figure}[!b]
    \centering
    \includegraphics[width=0.98\linewidth]{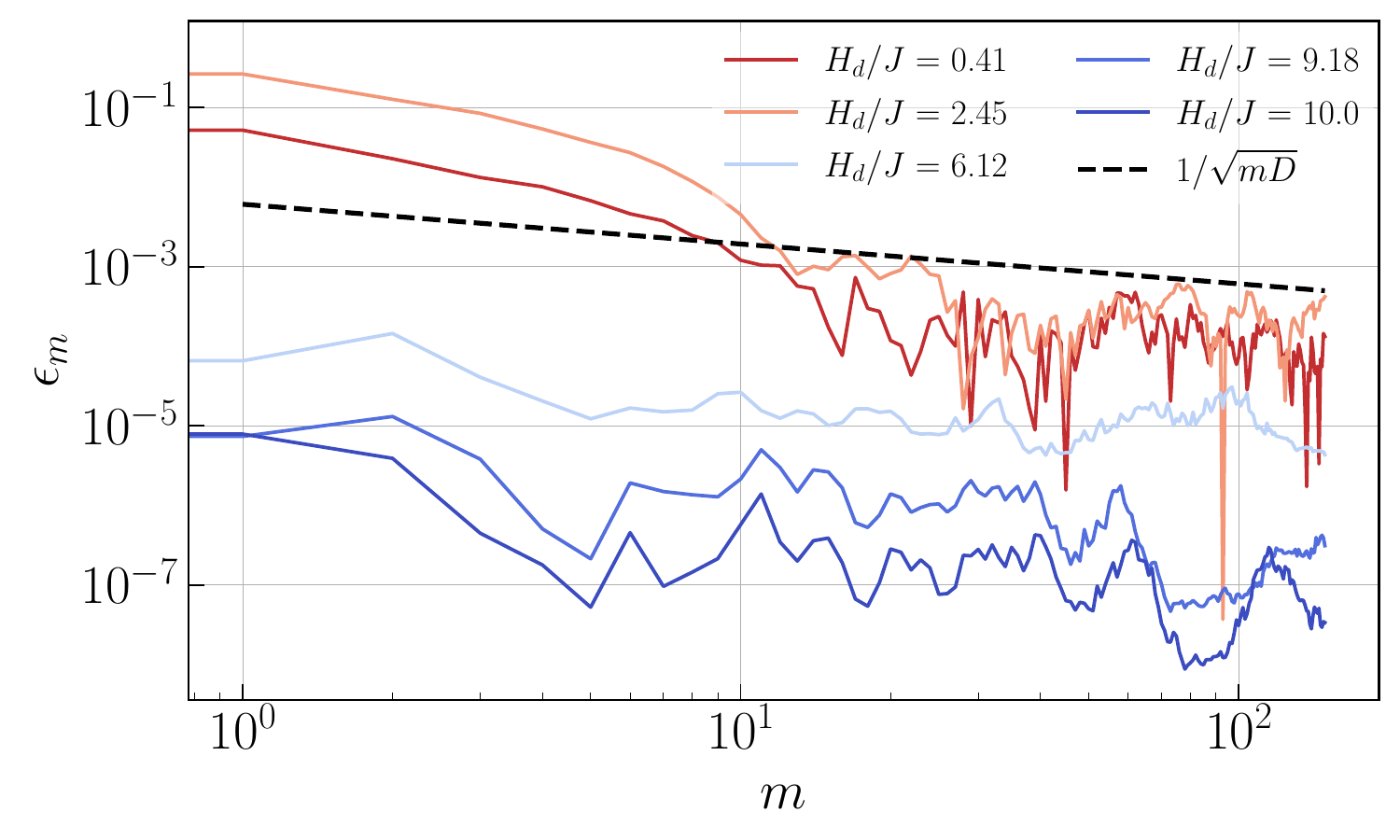}
    \caption{ The error $\epsilon_m$ of the DEA at each step during the first
       150 iterations of the algorithm.  
       Simulations were performed on a 20-site spin-chain, with red and blue
       lines indicating ergodic and dynamically frozen model parameters
       respectively. The dashed line shows a suggestive scaling for the error in the
       ergodic regime. }
    \label{fig:err}

\end{figure}

We analyze the DEA error $\epsilon_m$ at each iteration $m$ in \cref{fig:err}.
In the frozen regime $\epsilon_m$ quickly achieves a small value, $\epsilon_m
<10^{-6}$ within a few iterations, and decreases exponentially with increasing
drive strength. In the ergodic phase, the error rapidly decays to a scaling
$\epsilon_m \sim (mD)^{-1/2}$. This scaling arises due to the fact that the
initial state becomes delocalized over the Ritz vectors (which we later show
explicitly) and the DEA approximation becomes an average over $m$ matrix
elements each expected to obey ETH, which predicts fluctuations scaling as
$D^{1/2}$ for each matrix element~\cite{Rigol_Infinite_T}. Remarkably, this
indicates that the Ritz vectors obtained by diagonalizing the $m \times m$
projected Floquet unitary reproduce the fluctuations expected from the Floquet
eigenstates obtained by diagonalizing the exact $D \times D$ Floquet unitary. We
show in \cref{appendix:algorithm_choices} that this behavior is specific to the
presented algorithm and cannot be expected generically. Namely, alternative
diagonalization procedures such as an isometric Arnoldi iteration lead to long
tails and poor convergence in the ergodic phase. In the frozen regime, these
results indicate that the dynamics explores an effective Hilbert space which
does not grow with system size, and can be spanned to a very good approximation
by the Ritz vectors. Note that irrespective of the regime, the Ritz vectors
accurately describe the short-time dynamics \emph{by construction}: the Ritz
vectors in the Krylov subspace of dimension $m$ span the time-evolved states for
the first $m$ cycles by definition, since they are iteratively generated from
the initial state. As such, the first $m$ values of $\braket{\mathcal{O}(t)}$,
$t=0 \dots m-1$, are accurately described within this Krylov space, with the
only error appearing due to the loss of orthogonalization.
We here establish that the Ritz vectors also accurately capture the late-time
behavior.

\begin{figure*}[!ht]
    \centering
    \includegraphics[width=0.98\linewidth]{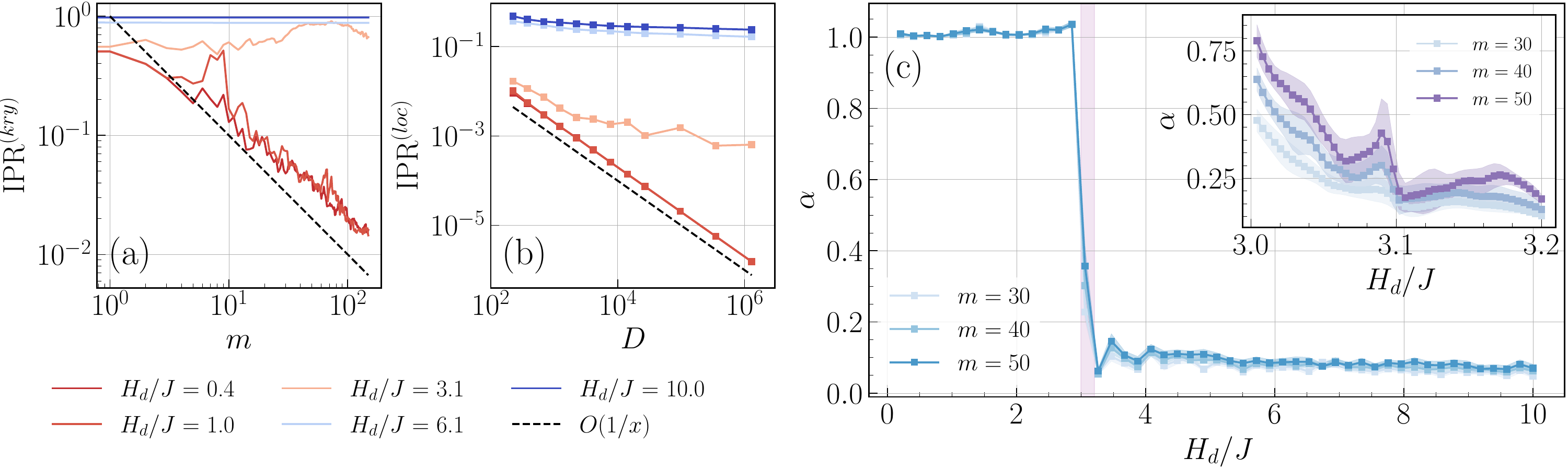}
    \caption{ Inverse participation ratios. (a) The inverse participation ratio
    of the initial state in the basis of Ritz vectors (approximate Floquet
    eigenvectors) for different Krylov space dimensions. Blue and red lines show
    frozen and ergodic model parameters respectively; the initial state can be
    seen to go between localized to delocalized in the Krylov subspace as the
    drive strength is increased. (b) The inverse participation ratio of the Ritz
    vectors in the computational basis, averaged over all Ritz vectors in a
    Krylov subspace of dimension 50. The dashed lines correspond to $1/m$ and $1/D$
    in (a) and (b) respectively. (c) We fit the lines from (b) to the form $\sim
    D^{-\alpha}$ and show the scaling exponent, with error bars, for different
    drive strengths $H_d/J$. Different lines correspond to different Krylov
    subspace dimensions, $m$. A sharp transition between localized and
    delocalized Ritz vectors in the full Hilbert space can be seen. The inset shows
    a more fine-grained parameter scan nearer the transition (purple shaded
    region of the main plot). }
    \label{fig:ipr}
\end{figure*}

\section{Localization of Ritz vectors}
\label{sec:ritzlocalozation}

We can make these statements more quantitative by analyzing the Ritz vectors and
the corresponding Krylov subspace. To this end, we compute the inverse
participation ratio (IPR)~\cite{wegner_inverse_1980,evers_fluctuations_2000} in
two complementary ways. The first is the IPR of the initial state with respect
to the Ritz vectors for a Krylov subspace of dimension $m$. We define
\begin{equation}
    \text{IPR}^{(kry)} = \sum_{i=0}^{m} | \langle \tilde{\mu}_i | \psi_0 \rangle |^4/\mathcal{N}^2
\end{equation}
which is inversely proportional to the number of Ritz vectors over which the
initial state is delocalized. A value close to one indicates a localized initial
state in the Krylov subspace, whereas a value of $1/m$ indicates complete
delocalization. $\text{IPR}^{(kry)}$ is shown in \cref{fig:ipr}(a) for different
drive strengths as the Krylov space dimension increases. The two dynamical
regimes are clearly distinguished by their scaling behaviors: in the ergodic
phase, the initial state is delocalized over the basis of Ritz vectors, whereas
in the dynamically frozen regime, only a small subset contributes significantly.

We also examine the localization of the individual Ritz vectors in the
computational basis, 
\begin{equation}
    \text{IPR}^{(loc)}_j = \sum_{i=1}^{D} | \langle c_i | \tilde{\mu}_j \rangle |^4,
\end{equation}
where $\{|c_i\rangle\}$ denotes the set of computational basis vectors. This
quantity measures how localized each Ritz vector is in the computational basis.
Averaging this IPR across all of the Ritz vectors for a given $\mathcal{K}_m$,
\begin{equation}
    \text{IPR}^{(loc)} =\frac{1}{m} \sum_{j=1}^{m} \text{IPR}^{(loc)}_j
\end{equation}
gives us a measure of how localized or delocalized the Krylov subspace is as a
whole. \Cref{fig:ipr}(b) shows this average for increasing Hilbert space
dimension with red and blue lines corresponding to ergodic and DF drive strength
parameters respectively. Again, the two phases demonstrate differing
localization properties. We observe a delocalized nature with $O(1/D)$ scaling
of  $\text{IPR}^{(loc)}$ in the ergodic regime and a localized nature in the
dynamically frozen regime. The delocalization in the ergodic regime corresponds
to the scaling expected from exact Floquet eigenstates.

This scaling was obtained by diagonalizing a smaller $m \times m$ matrix in a
Krylov basis that contains highly nonthermal states (i.e. the time-evolved state
at short times). On the level of the IPR, the Ritz vectors are statistically
indistinguishable from exact Floquet eigenstates. Fitting $\text{IPR}^{(loc)}$
to a power law reveals a sharp crossover between the delocalized and localized
regimes, as shown in \cref{fig:ipr}(c). Near the transition (inset), the scaling
deviates from the simple localized/delocalized limits, indicating
multifractality, and this behavior appears to be sensitive to the Krylov
subspace dimension $m$; we leave a detailed study of this dependence for future
work.

This analysis confirms that in the ergodic regime, the initial state explores a
large part of the Krylov subspace, whilst in the case of DF it remains frozen
and explores little of the generated subspace. The former is consistent with the
interpretation of ergodicity as delocalization in Krylov
space~\cite{balasubramanian_quantum_2022,scialchi_integrability--chaos_2024}.

\section{Discussion and Conclusion}
\label{sec:discussion}

We here illustrated the efficiency of the Krylov approach to the DEA in the
ergodic and dynamically frozen regime of a driven mixed-field Ising model. 
Beyond this specific application, we additionally show in \cref{appendix:fmbl}
that our algorithm can be used to efficiently compute the DEA for a Floquet
system in the many-body localized (MBL) regime, where the usual analysis with
exact diagonalization again severely limits the size of the systems that can be
studied. Taken together, these findings suggest the Ritz vectors constitute
a powerful diagnostic for characterizing ergodicity-breaking transitions, while
also providing a compact, reduced Hilbert space within which the system dynamics
can be accurately described across all time scales.

The efficiency of the Krylov approach can be directly related to the physics of
the different regimes. In the dynamically frozen regime, where one may expect
the time-evolved state to remain close to the initial state, it is natural to
expect that Krylov methods perform well. By applying the unitary Floquet
operator to the initial state, one quickly finds the relevant few eigenstates
accounting for a correction in the DEA on the trivial value given by the
initial state. Here, we essentially build a subspace that is highly non-thermal
and find that our initial state remains tightly localized within a small part of
this subspace. In the ergodic regime, however, one may expect that the
infinite-time average requires knowledge of all the Floquet eigenstates of the
system to accurately compute the DEA. Here, the Krylov subspace algorithm still
performs well, but for different reasons. In this regime, ergodicity actually
plays to our advantage. While the initial state indeed has overlap with many of
the Floquet eigenstates, the eigenstate thermalization hypothesis (ETH) dictates
that these eigenstates are themselves thermal, and typicality implies that one
can compute observables using a small subset of (approximate) eigenstates. We
find that our algorithm is capable of accurately generating representative
thermal states of the system within a number of iterations, and once the Krylov
space becomes large enough to reflect these states, the DEA essentially averages
over the expectation values of these thermal states; the Krylov subspace is
highly thermal and the initial state rapidly explores this subspace, as seen by
the respective IPRs. 

These results motivate the use of the Ritz vectors as an appropriate set of
dynamical eigenstates in both ergodic and nonergodic dynamics. Given a set of
such states, it is now possible to apply further probes of quantum chaos through
e.g. studies of their
entanglement~\cite{kumar_entanglement_2011,vidmar_entanglement_2017,haque_entanglement_2022,bianchi_volume-law_2022,kliczkowski_average_2023,herrmann_deviations_2025}
or fidelity
susceptibility~\cite{wang_fidelity_2015,pandey_adiabatic_2020,bhattacharjee_sharp_2024,abdelshafy_onset_2025},
or study the effect of conservation laws or kinetic constraints. This approach
can also be contrasted with MPS-based
methods~\cite{Cakan2021,haegeman_time-dependent_2011,paeckel_time-evolution_2019,kloss_studying_2020},
which exhibit entanglement barriers before thermalizing and do not provide a
`complete' eigenbasis in which the dynamics can be described. This work fits
within the broader topic of studying quantum chaos and ergodicity using Krylov
approaches~\cite{parker_universal_2019,balasubramanian_quantum_2022,rabinovici_krylov_2022,espanol_assessing_2023,erdmenger_universal_2023,scialchi_integrability--chaos_2024,rabinovici_krylov_2025,nandy_quantum_2025,suchsland_krylov_2023,baggioli_krylov_2025,balasubramanian_quantum_2025}.
Rather than focusing on the growth of and (de)localization within the Krylov
subspace, we here consider \emph{Floquet eigenstates} constructed in this Krylov
subspace. We show how these are able to capture early- and late-time dynamics in
both ergodic and frozen systems, and present a reduced basis through which
quantum many-body dynamics can be efficiently and systematically studied, here
exemplified through the DEA.

We have put forward an efficient Krylov subspace algorithm for computing the
infinite-time average of observables in periodically driven systems capable of
going beyond the size limitations of exact diagonalization. We have demonstrated
the effectiveness of the algorithm in computing the DEA of the total
magnetization density in a model exhibiting a transition to a dynamically frozen
regime. The nature of the Krylov subspaces has been characterized and it was
shown that in the ergodic phase, our subspace is highly thermal, while in the
dynamically frozen phase it quickly generates the relevant corrections to the
trivial DEA given by the initial state. Lastly, we have found that the algorithm
succeeds in computing the DEA for a Floquet-MBL system, and we expect it to
prove useful in a variety of other exciting settings relevant to the study of
Floquet matter.


\section*{Acknowledgments}
We thank Arnab Das for insightful discussions. A.W. acknowledges support by the
German Research Foundation (DFG) through the Emmy Noether program (Grant No.
509755282). A.H. was supported by the Marie Sk\l{}odowska-Curie grant agreement
No. 101110987. L.S., A.W., and P.W.C. acknowledge support from the Max Planck
Society and the computing resources at the Max Planck Institute for the Physics
of Complex Systems. The ED simulations were performed using the XDiag library
\cite{xdiag}. 

\bibliography{main}

\appendix
\section{End matter}


\subsection{Arnoldi iteration algorithm}
\label{appendix:algorithm}

To evaluate the DEA approximation $\braket{ \mathcal{O} }_{\mathrm{DEA},m}$
numerically we propose an algorithm employing the Arnoldi iteration
\cite{Arnoldi1951}. Given the initial vector $\ket{\psi_0}$ we iteratively
construct an orthonormal basis $\{\ket{v_k}\}_{k=1,\ldots,m}$ of the $m$-th
Krylov space $\mathcal{K}_m$ using the Arnoldi recursion, 
\begin{align}
\begin{split}
&\ket{v_1} = \ket{\psi_0} / \norm{\psi_{0}},\\
&\ket{w_k} = \mathcal{U} \ket{v_k},  \\
&\ket{\hat{v}_{k+1}} = \ket{w_k} - \sum\nolimits_{i=1}^{k} \braket{v_i | w_k}\ket{v_i},  \\
&\ket{v_{k+1}} = \ket{\hat{v}_{k+1}} / \norm{\hat{v}_{k+1}}, 
\end{split}
\label{eq:arnoldirecursion}
\end{align}
which can be understood as the successive Gram-Schmidt orthogonalization of the
spanning vectors $\mathcal{U}^{k}\ket{\psi_0}$. Combining the Arnoldi vectors
$\ket{v_k}$ as columns of the isometric matrix $V_m = \left( v_1 | \cdots |
v_{m} \right )$ we write the recursion \cref{eq:arnoldirecursion} as, 
\begin{equation}
    \label{eq:arnoldiprojection}
    \mathcal{U} V_m = V_{m+1} \hat{U}_{m+1}, 
\end{equation}
where we introduce a $(m+1) \times m$ matrix $\hat{U}_m$,
\begin{equation}
  \label{eq:hmatrix}
  \hat{U}_m =
  \begingroup 
  \setlength\arraycolsep{2pt}
  \begin{pmatrix}   
    u_{1,1} & u_{1,2} & u_{1,3} & \cdots & u_{1,m} \\
    u_{2,1} & u_{2,2} & u_{2,3} & \cdots & \vdots  \\
    0       & u_{3,2} & u_{3,3} &        & \vdots \\
    0       &       0 & \ddots & \ddots & u_{m-1,m} \\
    \vdots  &         &     0   & u_{m,m-1} & u_{m,m}\\
    0   &  \cdots     & 0       & 0      & u_{m+1,m}\\                     
  \end{pmatrix}
  \endgroup
  ,
\end{equation}
where $u_{i,j} = \braket{v_i| w_j}$. The  $m \times m$ submatrix $U_m$ obtained
from the first $m$ rows of $\hat{U}_m$ has zero elements $u_{i,j}$ whenever $i >
j+1$, and is commonly referred to as a Hessenberg matrix. The $i$-th eigenvalue
of $U_m$ is given by $\tilde{\xi}_i$ as in
\cref{eq:floquetoperatorprojectedeigenvalues}, whereas the corresponding Ritz
vectors are given by $\ket{\tilde{\mu}_{i}} = V_m \nu_i$, where $\nu_i \in
\mathbb{C}^m$ denotes the $i$-th eigenvector of $U_m$. Finally, the approximate
DEA $\braket{ \mathcal{O} }_{\mathrm{DEA},m}$ is computed by evaluating
\cref{eq:dea_krylov}. Notice that the definition of $\braket{ \mathcal{O}
}_{\mathrm{DEA},m}$ in \cref{eq:dea_krylov} is independent of the choice of
orthonormal basis of $\mathcal{K}_m$.

\subsection{Algorithmic Choices}
\label{appendix:algorithm_choices}

\begin{figure}[b!]
    \centering
    \includegraphics[width=0.98\linewidth]{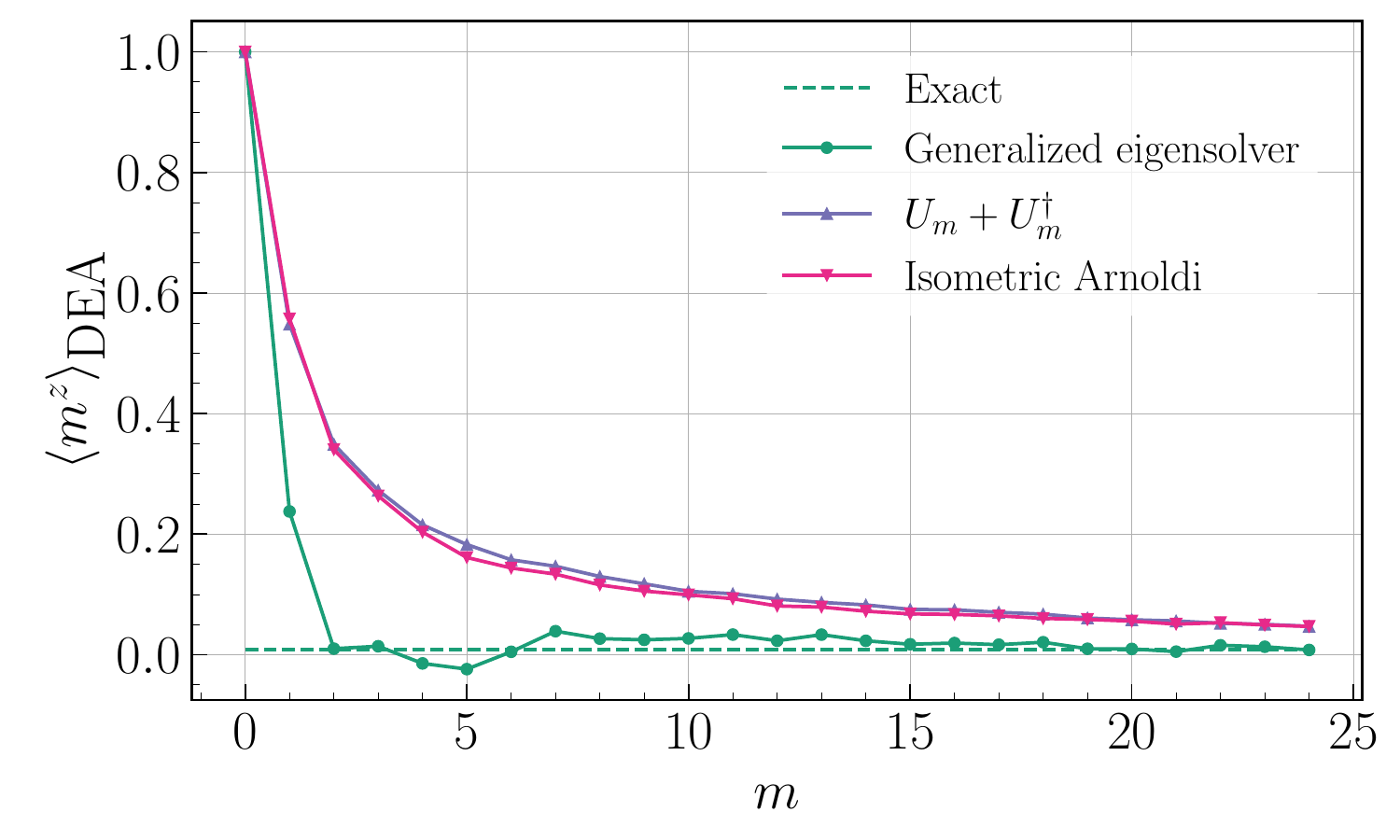}
    \caption{ We compare different numerical approaches to computing the DEA
    with our Krylov subspace algorithm. Dashed line denotes the exact DEA for
    the magnetization density on a spin-chain with 12 sites and $H_d/J=0.5$.
    Generalized eigensolver (green line) denotes the algorithm choice used in
    the main text where we take the right eigenvectors of $U_m$. Purple line
    shows the DEA computed with the eigenvectors of $U_m + U_m^\dagger$ and the
    pink line shows the DEA computed using an isometric Arnoldi procedure. }
    \label{app_fig:algorithm_choice}

\end{figure}

In \cref{appendix:algorithm}, we outlined an iterative algorithm for computing
the DEA of local observables. However, in projecting the unitary time evolution
onto the Krylov subspace, unitarity is lost. This means the corresponding matrix
$U_m$ is also no longer unitary and no longer shares the same eigenvectors as
$U_m + U_m^\dagger$. One way to remedy this is to rescale the final column of
the matrix $U_m$ in a procedure known as the isometric Arnoldi
iteration~\cite{arnoldi_isometric} to obtain a canonical unitary approximation to $U_m$. Another option is to compute the
eigenvectors of $U_m + U_m^\dagger$. Lastly, a
generalized eigensolver routine can be used to find the left and right eigenvectors of $U_m$. The last
approach will in general lead to a loss of orthonormality of the right eigenvectors. One can make use of the biorthogonality between the two sets of vectors, however, this does not
guarantee a real value of the DEA. Instead we work with just the right
eigenvectors and introduce a normalization to the DEA in
\cref{eq:dea_krylov_norm}. We compare each of these three approaches to
numerically computing the eigenvectors of the projected Floquet operator in
\cref{app_fig:algorithm_choice}. We show the same magnetization computed in the
main text for a spin-chain of 12 sites during the first 25 iterations of the
algorithm. We find that both the isometric Arnoldi routine as well as the
eigenvectors from $U_m + U_m^\dagger$ lead to long tails and poor convergence
for an ergodic choice of parameters, $H_d/J=0.5$. We see a notably faster
convergence to the ergodic DEA value (dashed line) when using the right
eigenvectors of $U_m$ and using the normalization in \cref{eq:dea_krylov_norm}
for the DEA as used in the main text. We note that in the DF frozen regime there
is no loss of orthonormality, the normalization $\mathcal{N}$ is unity and all
approaches give consistent results.

\subsection{Floquet-MBL}
\label{appendix:fmbl}

\begin{figure}[htb!]
    \centering
    \includegraphics[width=0.98\linewidth]{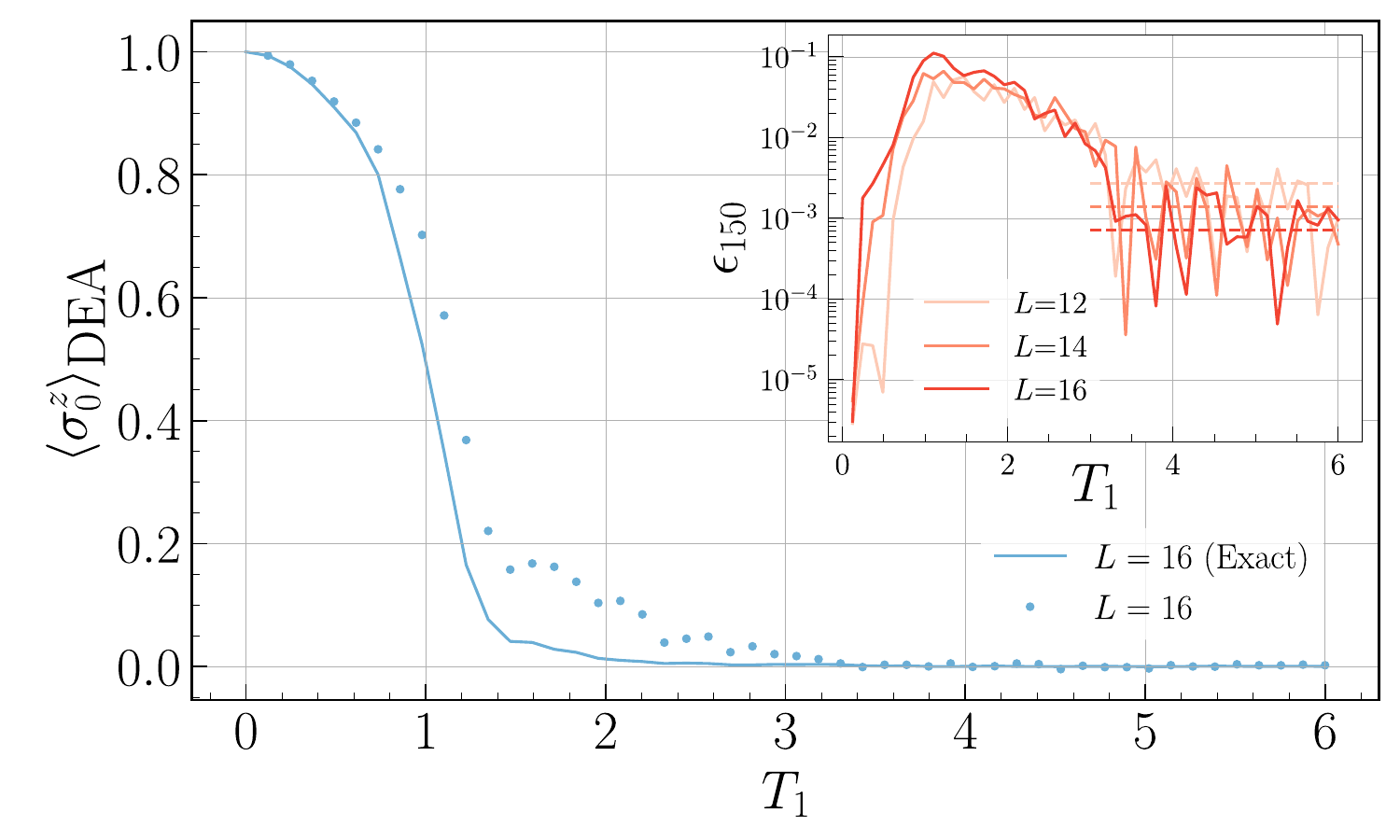}
    \caption{ Krylov subspace computation of DEA for MBL in a periodically
        driven system. We show the DEA for local magnetization, both exactly and
        following 150 iterations of our algorithm on a chain of 16 sites. The
        inset shows the corresponding error for several system sizes. }
    \label{app_fig:flMBL}

\end{figure}

In the main text, we employed our Krylov subspace algorithm for a system with
dynamical freezing. We now demonstrate its effectiveness for a periodically
driven system in the presence of disorder and compute the DEA of local
observables across an expected many-body localized transition. The following
periodic drive is prescribed,
\begin{equation}
    H(t) =  
    \begin{cases}
        J_x \sum_i \sigma^x_i\sigma^x_{i+1} + \sigma^y_i\sigma^y_{i+1} \quad  \text{for}  \quad 0 \le t \le T_1.\\
        \sum_i h_i \; \sigma_i^z + J_z \; \sigma^z_i \sigma^z_{i+1} \quad  \text{for} \quad T_1<t< (T_1 + T_0),  
    \end{cases}
    \label{Eq:H_Flq_2} 
 \end{equation} 
with period $T = T_0 + T_1$. The $h_i$ are chosen uniformly from an interval
$[-W,W]$, $J_x= J_z= \frac{1}{4}, T_0 =1, W=2.5$, as set out
in~\cite{Ponte_2015} with open boundaries. The former and latter parts of the
drive serve delocalizing and localizing purposes respectively. Starting from an
initial Néel product state $\ket{\psi_0}=\ket{\uparrow \downarrow
\uparrow\cdots}$  we compute the DEA of the local magnetization on a given site
as the delocalizing time $T_1$ is varied. \Cref{app_fig:flMBL} shows the
long-time average of the local magnetization computed both exactly and after
just 150 iterations of our algorithm for a chain of length 16 sites. The
algorithm performs best in the two limits of small $T_1$, where the eigenstates
are localized, and large $T_1$, where the phase is ergodic. The inset shows the
corresponding error for several system sizes. Similarly to the case of DF, we
observe a large compression of the error down to $10^{-5}$ in the localized
regime and an error that seems consistent with the scaling $\sqrt{mD}$ in the
ergodic regime. Whilst near the transition the convergence suffers, our algorithm
is nonetheless able to successfully capture a localized to delocalized
transition and allow for both the study of larger systems and further
diagnostics of the transition based on the knowledge of the Ritz vectors.

\end{document}